# HAXPES for materials science at the GALAXIES beamline

J.-P. Rueff[1,2,†], J. E. Rault[1], J. M. Ablett[1], Y. Utsumi[1*], and D. Céolin[1]

[1]Synchrotron SOLEIL, L'Orme des Merisiers, BP 48 St Aubin, 91192 Gif sur Yvette, France

[2]Sorbonne Université, CNRS, Laboratoire de Chimie Physique - Matière et Rayonnement, LCPMR, 75005 Paris, France

[†] rueff@synchrotron-soleil.fr

## 1 Introduction

The need for new materials is constantly growing as their use becomes more and more critical for technological advances. Materials come in a wide variety of structures, dimensionalities and phases. They can be functionalized or operated under the influence of various external parameters. This large diversity calls for new, powerful methods of characterization which are non-destructive and directly applicable to real materials or devices without the need of sample preparation and possibly under operando conditions. Hard x-ray photoelectron spectroscopy (HAXPES) is becoming one of the most effective methods of investigation of materials [1]. HAXPES combines the sensitivity of photoemission to the local chemistry and solid state effects with the 'ease of use' of a true bulk sensitive probe. HAXPES extends the application of standard x-ray photoemission to the domain of high-kinetic energy, therefore significantly augmenting the probing depth of the electrons to tens of nm. The recent surge of HAXPES has benefitted from the development of novel high performance photoelectron analyzers optimized for high kinetic energies (up to 10-15 keV) along with the advent of high brilliance x-ray sources available on the 3[rd] generation synchrotron facilities.

In this article, we will discuss the recent developments of HAXPES at the GALAXIES beamline at the SOLEIL synchrotron for the study of advanced materials.

## 2 HAXPES endstation

### 2.1 Overview

The GALAXIES beamline is a high flux undulator beamline dedicated to x-ray spectroscopy in the 2.3-12 keV photon energy with horizontal linear polarization. The beamline comprises of two endstations for inelastic x-ray scattering and HAXPES which are installed in two separate hutches. The beamline optical layout and details of the HAXPES endstation have been described elsewhere [2,3]. We briefly recall here their main characteristics.

The beamline x-ray energy is selected by a double-crystal, cryogenically cooled Si(111) monochromator (DCM) followed by a four-bounce high resolution monochromator (HRM) that is undergoing commissioning for 2019 operations. Currently, high energy resolution ($\Delta E \approx 150$ meV FWHM) can be obtained using the higher-order reflections above 6.9 keV. A first mirror serves to reject high-energy harmonics (using Pd and C selectable coatings) and to collimate the beam onto the focusing optics. For the HAXPES station, the beam is focused at the sample position using a Pd-coated toroidal mirror down to 30 µm (vertical) x 80 µm (horizontal) FWHM with a photon flux of the order of $10^{12}$ – $10^{13}$ ph / s using the first order of the DCM.

---

[*] now at Institute of Physics, Bijenička 46, 10000 Zagreb, Croatia

The HAXPES endstation operates up to 12 keV kinetic energy. The station layout is shown in Fig. 1(a). The station is equipped with a EW4000 SCIENTA hemispherical electron analyzer in the analysis chamber placed at 90° to the incident photon beam direction and along its horizontal electric field vector. The extra wide lens analyzer provides an enhanced collecting angle of photoelectrons which helps for compensating against the low photoionization cross-sections at high kinetic energies while maintaining an excellent resolution of the order 50 meV at 10 keV. The base pressure in the analysis chamber is around $2\times10^{-9}$ mbar and can be raised to $10^{-5}$ mbar for studies on diluted species and liquids without affecting the rest of the beamline vacuum thanks to the differentially pumped section (Fig. 1(a)) connected to the analysis chamber.

## 2.2  Surface preparation and sample environments

Samples can be mounted in the analysis chamber on a 4-axis cryogenically cooled (15 K), manipulator under UHV conditions. The sample transfer is realized through a loadlock (Fig. 1(b)) in the preparation chamber using a long linear transporter. The preparation chamber is equipped with an ion gun, evaporator, cleaving stage, annealing stage (up to 900° C) and LEED which permits basic sample preparation and surface characterization. Most of the time, the sample is loaded in the analysis chamber as-prepared without the need of any additional surface treatments. For a precise surface alignment, that is often required for angle-resolved measurements, the sample can be mounted on an additional holder equipped with a UHV piezo-driven rotary 'tilt' stage (Fig. 2(b)). The stage can be operated down to 50 K thanks to a thick copper braid connecting the sample holder to the cold finger.

We have developed a dedicated sample environment for in situ bias (Fig. 2(a)) using an ad-hoc sample holder and a metallic pad located on the back of the manipulator. The sample holder is equipped with a copper strip that can be grounded by contact with the metallic pad while the sample plate bias can be controlled by an external high voltage supply. This offers the possibility to carry out photoemission in operando on a variety of polarizable materials such a resistive switches, photovoltaic devices or materials for energy storage (cf. section 4.1). For air sensitive materials, a portable UHV suitcase developed with UPPA (Université de Pau et Pays de l'Adour) enables the transfer of samples directly from the glove box of the SOLEIL chemistry users laboratory to the loadlock (Fig. 1(b)).

More recently, a newly dedicated environment for liquid phase HAXPES studies has been commissioned and recently opened up to users. The setup shown in Fig. 3(a) consists of a microjet (Fig. 3(b)) designed by Microliquids that is incorporated into a differentially pumped UHV insert (Fig. 3(c)). The insert has two lateral openings for the photon entrance and exit and a skimmer on the analyzer side for photoelectron detection. The base pressure in the insert with the microjet running is maintained in the low $10^{-5}$ mbar range thanks to a LN$_2$ cold trap. The liquid microjet circulates with a HPLC pump between a 20 µm diameter glass capillary and a catcher which extracts the liquid (Fig. 3(b)) through a 300 µm hole. The alignment of the capillary, of the catcher and of the insert itself can be precisely controlled by motorized UHV actuators. A local heating stage permits to raise the temperature of the catcher and prevents frozen liquid from blocking it.

Finally, a gas cell from SCIENTA allows measurement of diluted species. The use of HAXPES in the gas phase falls beyond the coverage of this article but can be found in a recent review article [4].

## 3   HAXPES for depth profiling

HAXPES at GALAXIES has been used both for fundamental studies (e.g in semiconductors, oxides [5], topological insulator [6] or heavy fermion compounds [7]) and applied science. The large penetration depth of the electrons at high kinetic energy makes HAXPES an ideal tool for depth profiling. The effective probing depth $\lambda_{eff}$ can be varied according to:

$$\lambda_{eff} = \lambda \sin \theta$$

where $\lambda$ is the electron mean free path at the considered kinetic energy and $\theta$ is the electrons takeoff angle with respect to the sample surface. The effective probing depth $D$ is usually approximated by the empirical rule $D = 3\lambda_{eff}$. Surface and bulk sensitivity can be simply selected by varying the electrons takeoff angle from grazing to normal respectively. This method can be greatly improved by using the standing waves (SW) approach (cf section 3.1) in order to depth-resolve the relevant component contributions with atomic resolution.

Interestingly, the effective probing depth can be extended far beyond the $3\lambda$ rule by analyzing the inelastic photoelectric background using the Tougaard method [8]. In Ti/GaN for instance, the analysis was shown to provide depth distribution up to nearly 50 nm below the surface [9].

### 3.1   Depth-resolution via Photon Interferences Effects

a. Standing waves

Standing waves are generated by the interference between the incident and reflected x-ray photon fields generated near Bragg condition in a multilayer (ML) mirror or in a single crystal (Fig. 4(a)). SW photoemission consists of analyzing the photoelectron yield when excited by a standing wave. The nominal photon incidence angle $\theta_{inc}$ for forming the SW is defined by the first-order Bragg equation:

$$\lambda_X = d \cdot \sin \theta_{inc}$$

where $\lambda_X$ is the incident photon wavelength, d is the period of the ML mirror or the crystal d-spacing (cf. Figure 4(a)). The standing wave nodes and antinodes can then be moved through the sample depth by varying either $\theta_{inc}$ or $\lambda_X$ with a depth sensitivity of approximatively 1/10 of the SW period [10].

The interest of SW HAXPES is illustrated in the periodic (B$_4$C [10 Å]/ W [10 Å])$_{20}$ multilayer by computing the electric field intensity using the YXRO program [11] at 2.3 keV, the lowest energy accessible at the GALAXIES beamline (Fig. 4(b)). The wave field shows strong intensity modulation with a periodicity of 20 Å around the nominal angle $\theta_0 = 7.74°$ as given by the SW equation. These modulations allow the enhancement of photoelectron production at a well-defined depth in the materials depending on the choice of incidence angle (or photon energy). The E field intensity for two different incident angles around $\theta_0$ is shown as black lines in Fig. 4(b). We note a strong intensity modulation that can be chosen in phase or out of phase with respect to the ML structure. This method was for instance applied recently on the

GALAXIES beamline to characterizing the depth profile and inter-diffusion effects in Pd/Y ML X-ray mirrors [12]

Interestingly, the ML layer (or single crystal) can serve as a SW generator to characterize any aperiodic sample or monolayer deposited on the top of it. Fine tuning the incident angle (or energy) permits to characterize the depth profile of the deposited sample with atomic resolution. The depth profile of the sample can be reconstructed by modeling the materials profile and fitting the sample rocking curves, possibly in combination with x-ray reflectivity. This was applied recently at GALAXIES to characterize a graphene buffer layer grown on top of a SiC single crystal [13] or InAs nanoribbons transferred onto a Si/Mo ML [14].

### b. Near total reflection HAXPES

Often, complex or functional materials are aperiodic or cannot be easily deposited onto an adequate substrate for SW generation. It still is possible to generate a modulation of the electric field intensity which is significant enough for depth profiling, providing the optical densities of the different layers vary sufficiently across the sample depth. This approach was used at GALAXIES to determine the depth profile of $BiFeO_3$/$(Ca,Ce)MnO_3$ bilayer in a recent work by Marinova et al. [15]. The computation of the E field intensity shows poorly marked, yet strong modulation that can be used to retrieve the layer structure (Fig. 4(c)). The depth-resolved reconstructed profile shown in Fig. 4(d) is in excellent agreement with STEM-EELS profile. This allows a direct characterization of the polarization-induced charge density changes at the interface, providing insight on the ferroelectric switching effects

## 4 Operando conditions

### 4.1 In-situ bias

The study of solid-state electronic properties submitted to an external bias has gained spectacular importance recently with the massive development of a vast class of materials including ferroelectric and multiferroïc materials, memristive switches, or photovoltaic devices. Applying sample bias requires a nanometer thick metallic electrode to be deposited on top of the active area. Thanks to its high penetration capabilities and its chemical sensitivity, HAXPES makes possible to probe underneath the electrode, hence giving access to the electronic changes in operando as a function of applied bias.

On the GALAXIES beamline, in-situ bias has been successfully applied recently to ferroelectric PZT thin films [16] and organic switches [17]. In the PZT film, the application of an external bias lead to a change of the Pd and Zr core level binding energy (Fig. 5(b)) due to imperfect interface screening of the PZT ferroelectric polarization. An additional component due to the electrode-PZT interface is observed and is found detrimental to ferroelectricity.

An improved setup for in-situ bias is to be installed at GALAXIES during the second semester of 2018. The new setup will offer up to 13 contacts and extends the in-situ electrical connectivity to multiple leads or 4 probe resistivity measurements.

### 4.2 Materials for energy storage

There is a wealth of studies using HAXPES in materials for energy storage including materials for batteries, electrolytes and photovoltaics devices / solar cells. These benefit from the bulk sensitivity of HAXPES for probing real devices in operando or approaching realistic working conditions for instance in cathode materials. These different systems have been successfully investigated at GALAXIES. We refer the reader to the recent published works carried out for instance in superionic materials used as solid state electrolyte [18] and lithium rich cathodes for batteries [19].

### 4.3 Liquid phase

HAXPES in liquid phase has been recently opened to users at GALAXIES beamline. This permits to apply for the first time bulk sensitive electron spectroscopy to species in the liquid phase, thus avoiding surface sensitivity inherent to existing soft x-rays liquid setups. The first studies were performed on salts dissolved in water as they offer an ideal test systems for investigating charge transfer (CT) dynamics from solvent molecules to the solvated cations as exemplified in KCl [20].

The creation of 1s core-holes in $K^+$ and $Cl^-$ leads to very unstable species which relax in less than 1 fs mainly by $KL_{2,3}L_{2,3}$ Auger decay. In contrast to $Cl^-$, the presence of water molecules around $K^+$ ions leads to supplementary decay channels which were attributed to the ultrafast electron transfer from the water molecules to the unoccupied K 3d orbitals (Fig. 6). The strength of this CT mechanism depends on several factors such as the ionic charge, ion-water distances, electronic shell. This sensitivity opens up new perspectives for investigating ultrafast chemical dynamics in liquids.

## 5 Conclusions and perspectives

In this review, we have illustrated the interest of HAXPES for materials studies as performed at the GALAXIES beamline in various systems and sample environments including in-situ biasing or liquid phases. Several developments are underway which will add to the instrument portfolio for investigating the materials electronic properties. Hard x-ray angular resolved photoemission (HARPES) is currently under test at GALAXIES with the new tilt stage. This will extend the use of standard ARPES for resolving band structure to bulk materials, buried layers or interfaces [21]. We have developed in-situ magnetization capacity using a permanent magnet which can be used for magnetic-circular dichroism (MCD) HAXPES. Circular light is currently available at GALAXIES thanks to a quarter wave plate. MCD-HAXPES will enable the investigation of the magnetic properties of buried layers [22]. Finally, resonant photoemission or Auger will open up new perspective for enhancing orbital selectivity, or probing ultrafast charge dynamics through the core-hole clock method.

# Figures

**Figure 1**: The HAXPES endstation at the GALAXIES beamline, SOLEIL Synchrotron. (a) general view of the endstation showing the preparation chamber, the analysis chamber with the spherical electron analyzer and the sample main manipulator ; (b) close-up of the loadlock in the preparation chamber and the portable UHV suitcase.

**Figure 2**: Dedicated sample environment for in situ bias (a) and tilt stage (b)

**Figure 3**: Overview of liquid microjet setup at GALAXIES (a). The microjet head (b) is inserted into a differentially pumped tube (c) equipped with apertures for the photons and electrons. The head is composed of a glass capillary and a catcher. The liquid is injected using a HPLC pump and is extracted by the catcher which is permanently pumped.

**Figure 4**: (a) Geometry of the standing wave method on a multilayer sample; (b) computed electric field intensity generated by standing waves in B4C / W multilayers using the YXRO program. Black lines are intensity vs. depth cuts for two incidence angles ; (c) calculated electric field magnitude for the BFO/CCMO/STO heterostructures (d) ; (e) experimental (circles) vs theoretical (plain lines) rocking curves for the heterostructures. Thickness reported in (d) are obtained by fitting the theoretical curves to the experiment.

**Figure 5**: (a) Specific sample environment for in-situ bias ; (b) diagram of the device under investigation ; (c) bias-dependent core-levels of Pb, Zr and Ti showing ferroelectric-induced binding energy shifts due to imperfect interface screening.

**Figure 6:** KLL Auger spectra of $K^+$ (a) and $Cl^-$ (b) in aqueous KCl solution recorded at 5keV photon energy. The bars indicate the computed energies of different final states. All low-energy satellites in the theoretical spectra are charge transfer (CT) states attributed to electron transfer from water (W) to either $K^+$ (localized CT) or $Cl^-$ (delocalized CT).

**Figure 1**

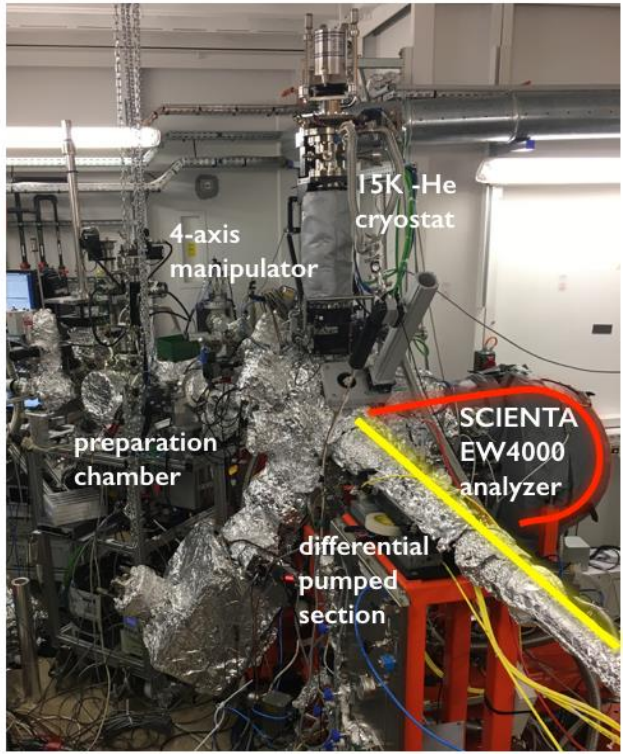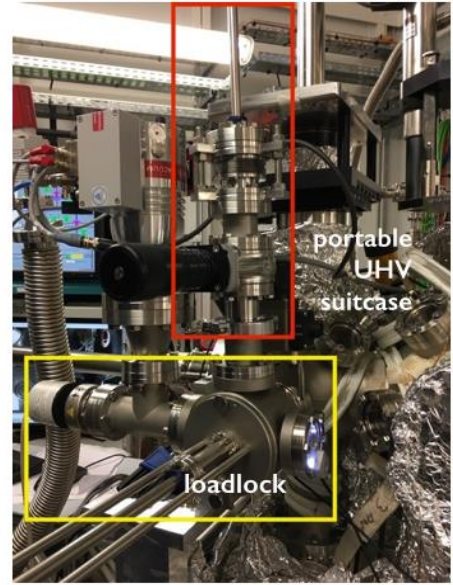

Figure 1: (a) Preparation chamber with 4-axis manipulator, 15K He cryostat, SCIENTA EW4000 analyzer, and differential pumped section. (b) Portable UHV suitcase and loadlock.

**Figure 2**

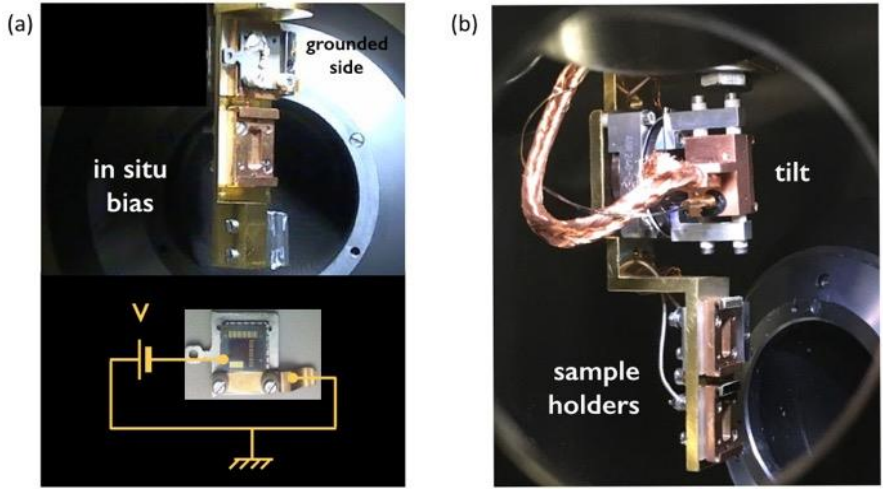

**Figure 3**

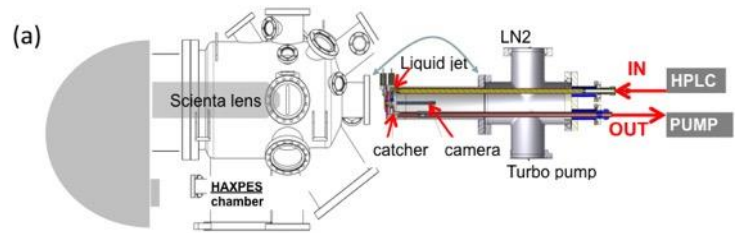
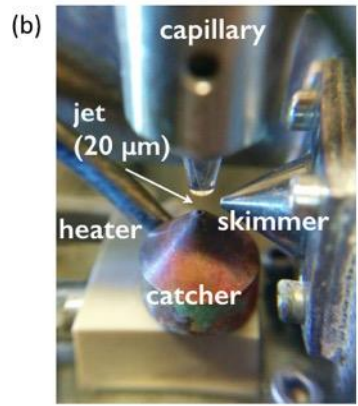
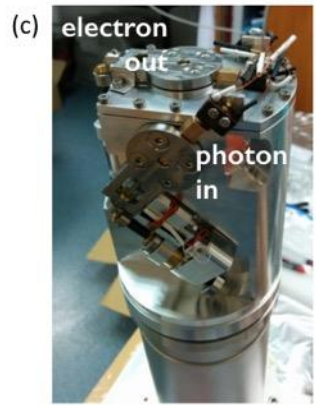

**Figure 4** :

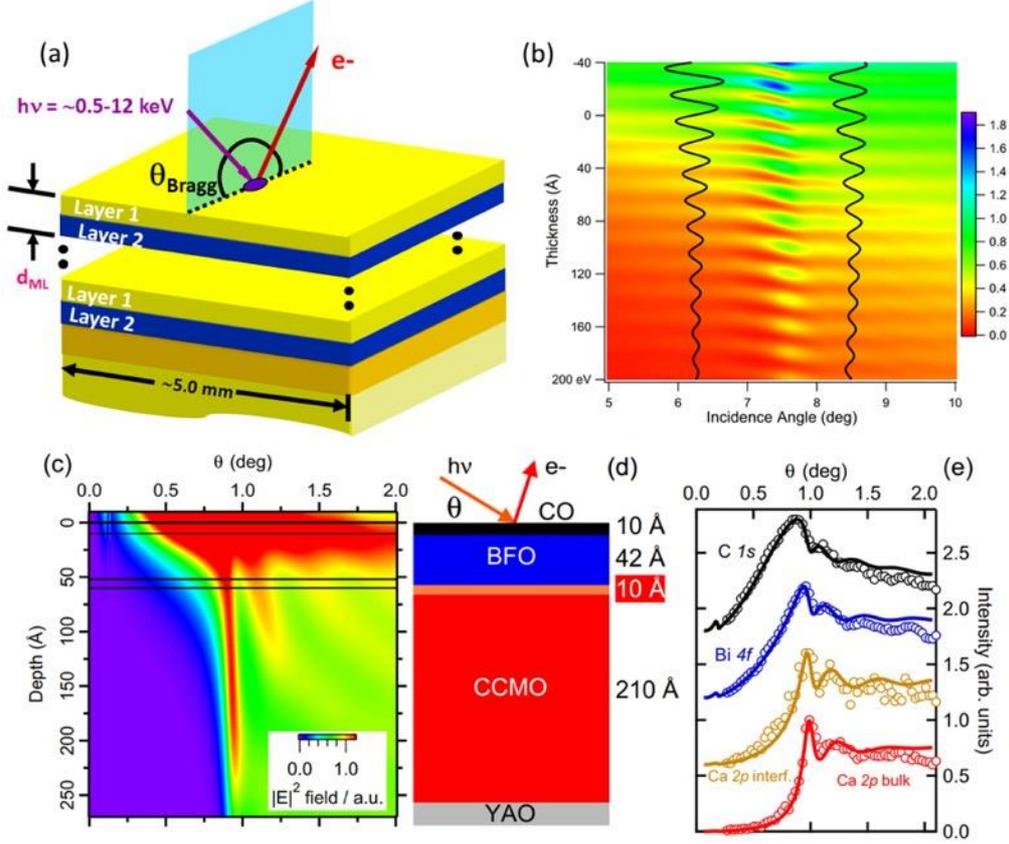

**Figure 5** :

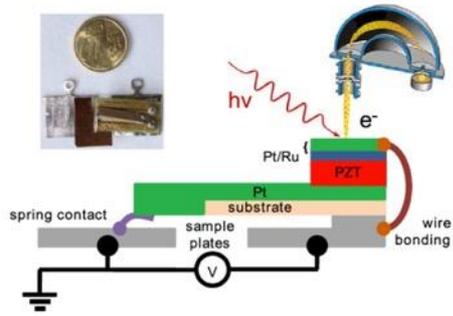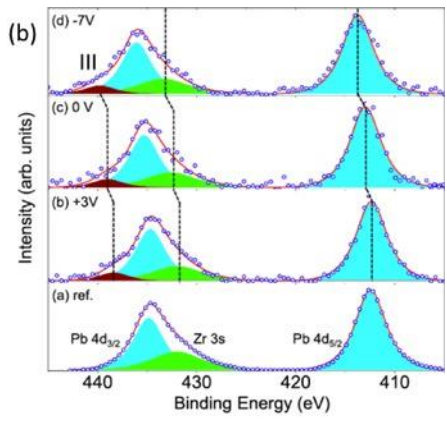

**Figure 6**

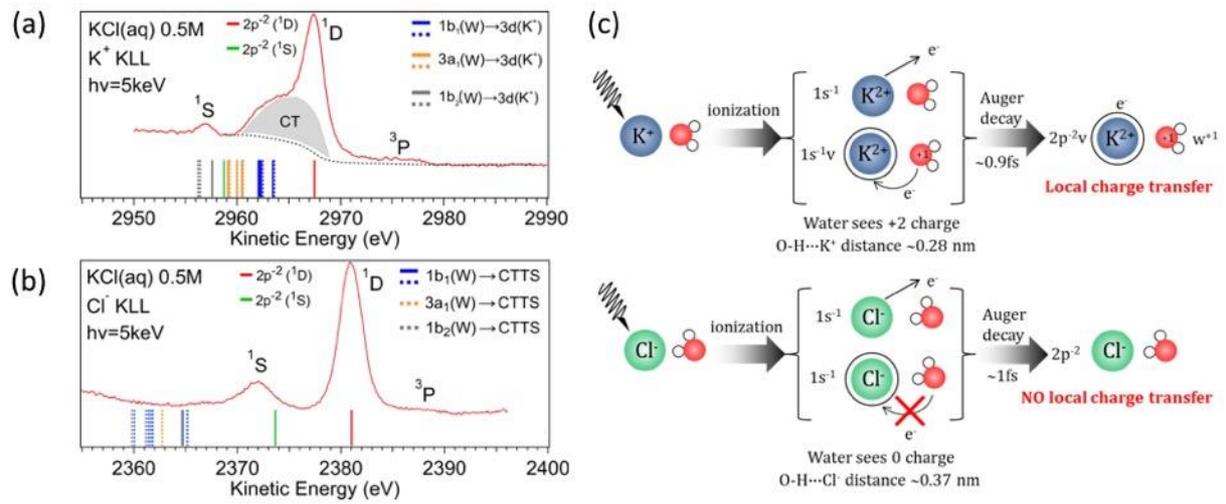